\documentclass[prb,floatfix,showpacs,superscriptaddress,longbibliography,twocolumn]{revtex4-2}

\usepackage[english]{babel}
\usepackage{amsmath,amssymb,amsfonts}
\usepackage{epsfig}
\usepackage{graphicx}
\usepackage{outlines}
\usepackage[dvipsnames]{xcolor}
\usepackage{changes}
\usepackage[colorlinks=true,linkcolor=blue,citecolor=red]{hyperref}

\DeclareMathAlphabet\mathbfcal{OMS}{cmsy}{b}{n}

\newcommand{\figu}[1]
{Fig.~\ref{#1}}

\def\a{\alpha}       \def\b{\beta}

                    \def\s{\sigma}
\def\t{\tau}

\def\PP{{\cal P}} 

\def\TT{{\cal T}}

\def\=={\equiv}

\def\iome{i\omega_n}

\def\up{\uparrow} \def\down{\downarrow} \def\dw{\downarrow}

\def\ka{{\bf k}}
\def\ia{{\bf r}}

\def\qa{{\bf q}}

  \def\Tr{{\rm Tr}\,}

\def\bra{\langle} \def\ket{\rangle}

\usepackage{bbold}
\def\11{\mathbb{1}}
\def\00{\mathbf{0}}

\newenvironment{eqs}%
{\begin{equation} \begin{aligned}}%
{\end{aligned} \end{equation} }
\newcommand{\beal}{\begin{eqs}}
\newcommand{\eal}{\end{eqs}}

\newcommand{\bealn}{\beal\nonumber}
\newcommand{\dagga}{{\phantom{\dagger}}}
\newcommand{\fract}[2]{\frac{\displaystyle \;#1\;}{\displaystyle \;#2\;}}
\newcommand{\eqn}[1]{(\ref{#1})}
\newcommand{\mket}[1]{| #1\rangle}

\begin{document}
\title{ Exciton condensation in strongly correlated quantum spin Hall insulators}

\author{A.~Amaricci}
\affiliation{CNR-IOM, Istituto Officina dei Materiali,
Consiglio Nazionale delle Ricerche, Via Bonomea 265, 34136 Trieste, Italy}

\author{G.~Mazza}
\affiliation{Dipartimento di Fisica, Universit\`a di Pisa, Largo Bruno Pontecorvo 3, 56127, Pisa, Italy}
\affiliation{Department of Quantum Matter Physics, University of
  Geneva, Quai Ernest-Ansermet 24, 1211 Geneva, Switzerland}

\author{M.~Capone}
\affiliation{Scuola Internazionale Superiore di Studi Avanzati (SISSA), Via Bonomea 265, 34136 Trieste, Italy}
\affiliation{CNR-IOM, Istituto Officina dei Materiali,
Consiglio Nazionale delle Ricerche, Via Bonomea 265, 34136 Trieste,
Italy}

\author{M.~Fabrizio}
\affiliation{Scuola Internazionale Superiore di Studi Avanzati (SISSA), Via Bonomea 265, 34136 Trieste, Italy}

\begin{abstract}
Time reversal symmetric topological insulators are generically robust
with respect to weak local interaction, unless symmetry
breaking transitions take place.
Using dynamical mean-field theory we solve an interacting model of
quantum spin Hall insulators and show the existence, at intermediate
coupling, of a symmetry breaking transition to a non-topological
insulator characterised by exciton condensation. This transition is of
first order. 
For a larger interaction strength the insulator evolves into a Mott one.
The transition is continuous if magnetic order is prevented, and notably, for
any finite Hund's exchange  it progresses through a Mott localization
before the condensate coherence is lost.
We show that the correlated excitonic state corresponds to a
magneto-electric insulator which allows for direct experimental
probing. Finally, we discuss the fate of the helical edge modes across
the excitonic transition. 
\end{abstract}

\maketitle

The concept of  symmetry protected topology has introduced a new
paradigm for the description of electronic band structures \cite{Qi2011RMP,Wen2017RMP}.  
The early identification of topological states in 
semi-conducting quantum wells \cite{Bernevig2006S,Konig2007S} and 
three-dimensional chalcogenides \cite{HsiehN2008,HsiehS2009,Zhang2009NP,ChenS2009}
boosted intense research activity that finally reached a
mature symmetry groups classification for weakly-interacting
insulators and semi-metals. 
The discovery of topological properties in more correlated
materials \cite{Kargarian2013PRL,Herbut2014PRL}, such as 
monolayers of early transition-metals dichalcogenides (TMD)s
\cite{Qian2014S,Sun2022NP,Jia2022NP} or some Fe-based compounds
\cite{Hao2014PRX,Hao2015PRB,Wang2015PRB,Wang2018S,Chen2019SR}, raised interest in the role of the ever-present electron-electron interaction in topological
phases of matter.

The electron localization tendency brought in by strong correlations
can generically lead to dramatic modifications of the band structure topology \cite{Rachel2018ROPIP}.
Contrary to na\"{\i}ve expectations, Coulomb repulsion can in some cases favour the formation of a
non-trivial electronic state \cite{Zhang2009PRB,Herbut2014PRL}, trigger the existence of novel purely
interacting topological phases \cite{Pesin2010NP}, or drive a dynamical
change in the thermodynamic character of the topological quantum phase
transition \cite{Amaricci2015PRL,Roy2016,Amaricci2016PRB,Weyl2020PRR}. 
Yet, the most impactful effect of strong electronic correlations is
often the emergence of ordered phases.
At strong coupling, the existence of large spin-exchanges and spin-orbit coupling paves the way to magnetically ordered states.
For weaker interaction strength, the situation can get more intriguing since 
diverse degrees of freedom are equally
active and possibly cooperate with the non-trivial topology of
the electronic bands.
In these conditions, different instabilities compete and it becomes
hard to predict the electronic properties of a correlated topological insulator. 
 
One of the most interesting effect of electronic interaction in systems
hosting a small energy gap is to induce in-gap 
excitons \cite{Budich2014,Kunes2015JOPCM,Geffroy2019PRL,Mazza2020PRL,Blason2020PRB,Windgatter2021NCM}. 
Although excitons have been studied for long, recent
evidences supporting the existence of excitonic phases     
in TMDs mono-layers \cite{Qian2014S,Varsano2020NN,Sun2022NP,Jia2022NP}
gave strong impulse to the investigation of excitons in topological insulators~\cite{Budich2014,Geffroy2019PRL,Blason2020PRB}.
For instance, the anomalies observed in the topological Kondo insulator $SmB_6$ have been predicted to be caused just by 
excitons \cite{Dzero2010PRL,Dzero2012PRB,Zhang2013PRX,xLu2013PRL,Knolle2017PRL}. 

Here, we show that exciton phase transition generically occurs due to  
electronic correlations in a model quantum spin Hall
insulator (QSHI). In particular, using a non-perturbative approach based on
Dynamical Mean-Field Theory
(DMFT) \cite{GeorgesKotliar1996,metzvol,MullerHartmann1989},
we demonstrate that, in presence of a sufficiently strong 
interaction, the QSHI becomes unstable towards an excitonic
phase with an in-plane spin polarisation \cite{Geffroy2018PRB,Geffroy2019PRL} that  breaks the time-reversal, spin $U(1)$ and parity symmetries \cite{Blason2020PRB} that protect topological order. The transition between the QSHI and the  
Excitonic Insulator (EI) is of first order within DMFT.
The excitonic phase 
shows a finite magneto-electric susceptibility \cite{Vaz2010AM,Thole2020JOAP},
which allows a direct  experimental identification of such state of matter.  

The rest of the paper is organized as follows. In the Sec.~\ref{Sec1}
we introduce the interacting QSHI model and briefly recall the method
used to solve it. In the following section \ref{Sec2} we discuss the
excitonic phase transitions occurring for generic values of the
parameters, distinguishing the two cases corresponding to the presence or
the absence of Hund's exchange. We summarize part of the findings in
terms of phase-diagram in Sec.~\ref{Sec3}. In the section \ref{Sec4}
we discuss observables consequences of the excitonic transition in the
QSHI. Finally, in Sec.~\ref{Sec5} we draw the conclusions of
our work and discuss some perspectives.

\section{Model and Methods}\label{Sec1}
We consider an interacting two-orbital Hubbard model on a  two
dimensional square lattice \cite{Bernevig2006S,Amaricci2015PRL}, described by the Hamiltonian 
\beal
H=\sum_\ka\, \psi^\dagger_\ka\, H(\ka)\,\psi^\dagga_\ka + H_{int}\,,
\label{Ham}
\eal
with the spinor  $\psi^\dagger_\ka=\left[ c^\dagger_{\ka 1\up},\,
  c^\dagger_{\ka 2\up},\, c^\dagger_{\ka1\down},\, c^\dagger_{\ka  2\down} \right]$
and where $c^\dagger_{\ka\a\s}$ creates an electron on 
orbital $\a=1,2$, with spin $\s=\up,\dw$ at momentum $\ka$. 
Orbital 1 and 2 transform as the $\ell=0$ and $\ell=1$ spherical harmonics, respectively \cite{Bernevig2006S}, and, more specifically, 
\bealn 
(2,\up) &\equiv (\ell=1,\ell_z=+1,\up)\,,\\
(2,\down) &\equiv (\ell=1,\ell_z=-1,\down)\,,
\eal
are the $j_z=\pm 3/2$ components of $j=3/2$ spin-orbit multiplet.

We introduce the 
$4\times 4$ matrix basis $\Gamma_{\a a}=\sigma_\a
\otimes \tau_a$,  where $\s_{\a=0,1,2,3}$ and $\t_{a=0,1,2,3}$ are Pauli
matrices, including the identity, in spin and orbital subspaces, respectively. The non-interacting Hamiltonian matrix reads
\begin{equation}
H(\ka)=M(\ka)\,\Gamma_{03} +
\lambda\,\sin(k_x)\,\Gamma_{31}-\lambda\,\sin(k_y)\,\Gamma_{02}\,,
\label{Hbhz}
\end{equation}
where $M(\ka)= M-\epsilon\big(\cos k_x +\cos k_y)$, $M\geq 0$ being the energy separation between the two orbitals, $\epsilon$ the hopping amplitude
and $\lambda$ the inter-orbital hybridization that lacks an on-site component 
because of inversion symmetry. 
Hereafter, we take $\epsilon=1$ as our unit of energy, $\lambda=0.3$, and assume two electrons per site, i.e. half-filling.
The non-interacting Hamiltonian is invariant under Time Reversal
Symmetry $\TT$ (TRS), inversion symmetry $\PP$, $U(1)$ spin rotations around the $z$-axis, and the fourfold 
$C_4$ spatial rotations around $z$. 
We assume that the interaction is also invariant under the same symmetries, and, in particular, we take 
\begin{equation}
H_{int}=  \fract{1}{4}\,\sum_\ia
\big(2U\!-\!3J\big)\,\hat{N}_\ia^2 -
J\,\hat{S}_{z\ia}^2 + 2J\,\hat{T}_{z\ia}^2\,,
\label{Hint}
\end{equation}
where $\ia$ labels the sites of the lattice, and the operators
\bealn
\hat{N}_\ia &=\psi^\dag_{\ia}\,\Gamma_{00}\,\psi^\dagga_{\ia}\,,\\
\hat{S}_{z\ia} &= \fract{1}{2}\,\psi^\dag_{\ia}\,\Gamma_{30}\,\psi^\dagga_{\ia}\,,\\
\hat{T}_{z\ia}&=\fract{1}{2}\,\psi^\dag_{\ia}\,\Gamma_{03}\,\psi^\dagga_{\ia}\,,
\eal 
with $\psi_\ia$ the Fourier transform of $\psi_\ka$, 
are, respectively, the density, the spin polarization along $z$ and
the orbital polarization at site $\ia$. The interaction \eqn{Hint}
is not the most general symmetry allowed one. However, 
it enforces the first Hund's rule of maximum spin, whose role we aim
to analyse here.

We treat the interaction non-perturbatively by single-site
DMFT using exact
diagonalization as impurity solver \cite{Amaricci2022CPC}.
Within DMFT, the self-energy is approximated by a momentum independent but frequency dependent matrix function in spin and orbital space. 
A symmetry invariant self-energy matrix is diagonal, with spin-independent elements. Deviations from such matrix structure signal the onset of symmetry
breaking \cite{miyakoshiPRB87,Wang2012EEL,Yoshida2012PRB,Budich2012PRB,Amaricci2016PRB,Amaricci2018PRB,Geffroy2018PRB,Geffroy2019PRL}.
The non-interacting model has a topological quantum phase
transition between a QSHI for $M < 2$ and a trivial Band Insulator
(BI) for $M>2$.
In the presence of a finite Hund's exchange $J$ and for large $U$, see Eq.~\eqn{Hint}, a high-spin Mott insulator 
sets in \cite{Werner2007PRL,Budich2012PRB,Amaricci2015PRL} and 
describes two electrons localized on each site and forming a spin $S_z=\pm 1$ configuration, thus with vanishing orbital polarization $T_z=0$.

\begin{figure}
  \includegraphics[width=0.49\textwidth]{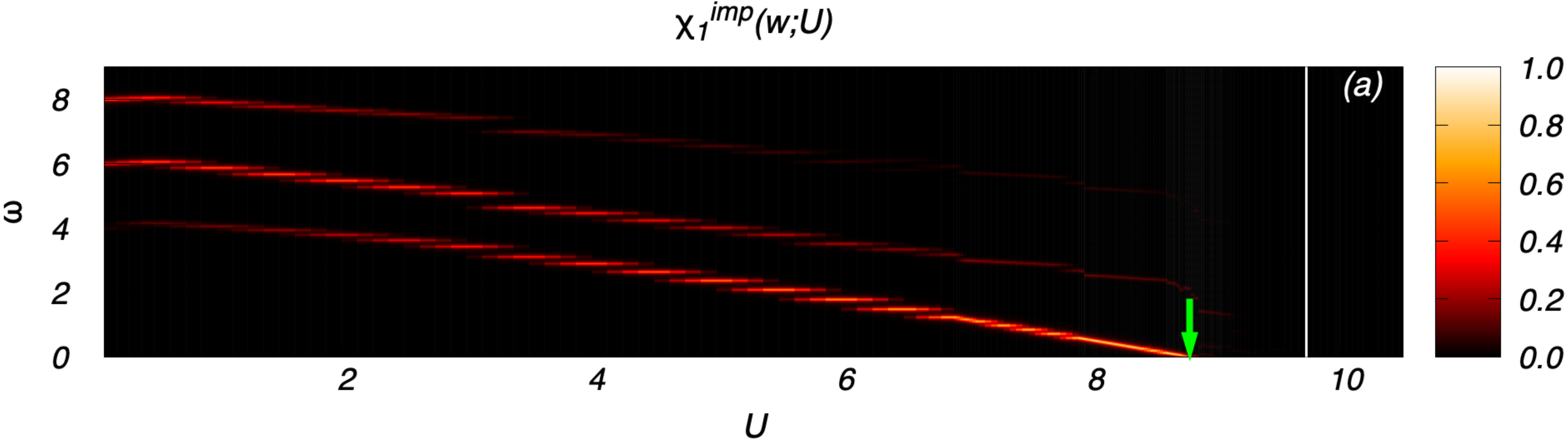}
  \caption{
    Evolution of the low-energy spectra of the in-plane triplet
    component of the exciton-exciton susceptibility
    $\chi^{imp}_1(\omega)$ as a function of the interaction strength $U$.    
    Data for $J/U=0.25$ and $M=3.5$. The arrow indicate the softening of the
    lowest energy peak before Mott insulator sets in (white solid line).
  }
  \label{fig0}
\end{figure}

\section{The Excitonic Phase-Transition}\label{Sec2}
In order to asses the possible instability of the model towards an excitonic
phase it is instructive to start from the atomic limit with two
electrons per site.
The Hamiltonian in the two-electron subspace reads 
\bealn
H_\text{at} &= \sum_\ia -J\,\hat{S}_{z\ia}^2 + 2J\,\hat{T}_{z\ia}^2 + 2M\,\hat{T}_{z\ia}\,.
\eal
The eigenstates can be labelled by the eigenvalues $S_{z}$, $T_{z}$ and $\ell_{z}$, respectively, of the  operators $\hat{S}_{z\ia}$, $\hat{T}_{z\ia}$ and 
\bealn
\hat{\ell}_z &= n_{2\up\ia}-n_{2\down\ia}\,,
\eal
with $n_{\a\sigma\ia}=c^\dagger_{\a\sigma\ia}c_{\a\sigma\ia}$. Thus
the states $\mket{(\ell_z,S_z,T_z),\ia}$ have eigenvalues 
$E\left(\ell_z,S_z,T_z\right)$:    
\beal
E(0,0,+1) &= 2J+2M  \,,\\
E(0,0,-1) &= 2J-2M  \,,\\
E(+1,+1,0) &= E(-1,-1,0) = -J  \,,\\
E(+1,0,0) &= E(-1,0,0) = 0\,.
\label{energies atomic limit}
\eal
For $3J>2M$ the atomic ground state is the high-spin doublet with $S_z=\pm 1$, 
otherwise is the state $\mket{(0,0,-1),\ia}$ with two electrons in orbital 2. 
Our aim is to study the competition between those states, and
therefore we hereafter drop the other three states, $\mket{(0,0,+1),\ia}$
and $\mket{(\pm 1,0,0),\ia}$.

Moreover, we define a pseudo spin operator $\mathbf{I}_\ia = \left(I_{x\ia},I_{y\ia},I_{z\ia}\right)$ through  
\bealn
I_{z,\ia}\mket{(+1,+1,0),\ia} &\equiv I_z\mket{+1,\ia} = \mket{+1,\ia}\,,\\
I_{z,\ia}\mket{(0,0,-1),\ia} &\equiv I_z\mket{0,\ia} = 0\,,\\
I_{z,\ia}\mket{(-1,-1,0),\ia} &\equiv I_z\mket{-1,\ia} = -\mket{-1,\ia}\,,
\eal
so that the three states become the components of an $I=1$ pseudo spin. 
In this subspace the following equivalences hold
\bealn
\psi^\dagger_\ia\Gamma_{11}\psi_\ia &\equiv \sqrt{2}I_{x\ia},&
\psi^\dagger_\ia\Gamma_{21}\psi_\ia &\equiv \sqrt{2}I_{y\ia},\\
\psi^\dagger_\ia\Gamma_{12}\psi_\ia &\equiv -\sqrt{2}\left\{I_{y\ia},I_{z\ia}\right\},&
\psi^\dagger_\ia\Gamma_{22}\psi_\ia &\equiv -\sqrt{2}\left\{I_{x\ia},I_{z\ia}\right\},\\
\psi^\dagger_\ia\Gamma_{03}\psi_\ia &\equiv 2\left(1-I_{z\ia}^2\right),&
\psi^\dagger_\ia\Gamma_{30}\psi_\ia &\equiv 2I_{z\ia},\\
\eal
while $\psi_\ia^\dagger\Gamma_{\alpha a}\psi_\ia$ with all other $\Gamma$ matrices 
different from the identity have vanishing matrix elements.

The atomic Hamiltonian projected onto the subspace 
$\mket{0,\ia}$ and $\mket{\pm 1,\ia}$  becomes, dropping constants:  
\bealn
H_\text{at} &\simeq \Delta E\,\sum_\ia\,\Big(1-I_{z\ia}^2\Big)\,,& 
\Delta E &= 3J - 2M\,.
\eal
Our interest is studying how the hopping processes beyond the atomic limit 
modify the level crossing between $\mket{\pm 1,\ia}$ and $\mket{0,\ia}$ when 
$\Delta E$ changes sign. For that, we treat those processes at second order in perturbation theory and, after projection onto the above subspace, we find an  effective Heisenberg Hamiltonian for the $I=1$ pseudo spins 
\beal
H_* &= \Delta E_*\,\sum_\ia\,\Big(1-I_{z\ia}^2\Big) \\
& +J_+\!\!\sum_{<\ia\ia'>}\Big(
2\,I_{z\ia}\,I_{z\ia'} - \sum_{a=x,y}I_{a\ia}\,I_{a\ia'}\Big)\\
& +J_-\!\!\sum_{<\ia\ia'>}\Big(
K_{z\ia}\,K_{z\ia'} - \sum_{a=x,y} K_{a\ia}\,K_{a\ia'}\Big)
\,,
\label{Ham-Heisenberg}
\eal
where $K_{a\ia} = \big\{ I_{a\ia},I_{z\ia}\big\}$, while 
\bealn
\Delta E_* &= \Delta E + 8J_-\,,&
J_\pm &= \fract{1\pm\lambda^2}{4U}\;.
\eal
When $\Delta E_*\gg J_+>J_-$, the ground state is a N\'eel antiferromagnet 
with $\langle\,I_{z\ia}\,\rangle=(-1)^\ia$. 
On the contrary, when $\Delta E_*\ll -\left|J_+\right|$,   
each site in the ground state is locked into the $I_z=0$ eigenstate of the pseudo-spin triplet, which is just the trivial band insulator since the topological one 
does not survive in the atomic limit. These two states 
might cross in energy when $\Delta E_* \simeq 0$, but that crossing is 
preempted by the quantum fluctuations brought about by $J_+$ and $J_-$ that, in turn, compete against each other. Since $J_+$ is larger, we can safely neglect 
$J_-$. In that case, the Hamiltonian \eqn{Ham-Heisenberg} describes an easy-axis spin-1 Heisenberg antiferromagnet with a single-ion anisotropy $\Delta E_*$, which suggests that the transition between the N\'eel antiferromagnet and the band insulator might occur through an intermediate phase characterised by the order parameter 
\beal
\Delta(\phi) &=
\langle I_{x\ia}+ I_{y\ia} \rangle
=
\frac{1}{\sqrt{2}}\langle\psi^\dagger_\ia \left( \Gamma_{11}+\Gamma_{21}\right)\psi^\dagga_\ia\rangle\\
&\equiv
\Delta_{11} + \Delta_{21} = \Delta\cos(\phi) + \Delta\sin(\phi)
\label{order parameter}
\eal
which breaks $\TT$, inversion symmetry $\PP$ and spin $U(1)$ symmetry for
any fixed value of $\phi\in[0,2\pi)$~\cite{Blason2020PRB}.
This phase actually describes a condensate of odd-parity spin-triplet excitons, 
with the spin lying in the $x-y$ plane. 

\begin{figure}
  \includegraphics[width=0.5\textwidth]{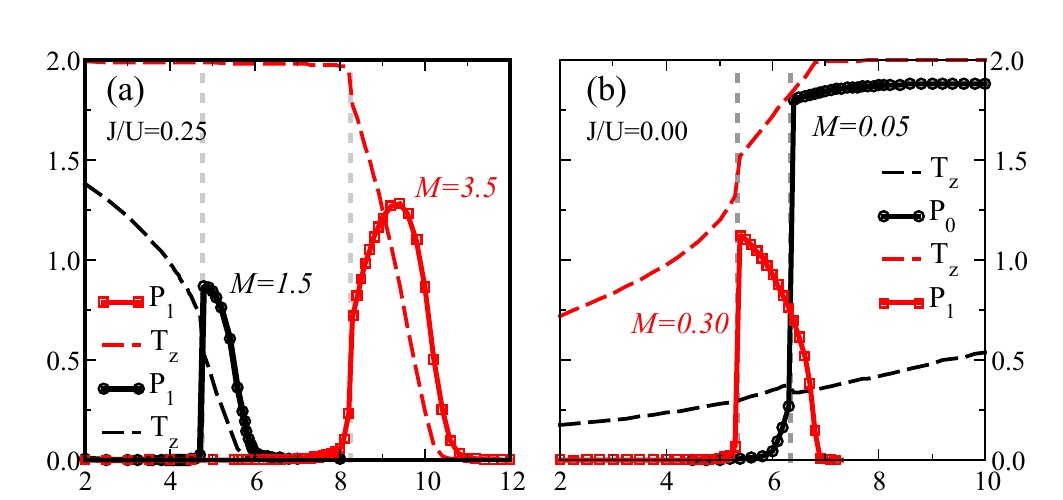}
  \caption{
    Orbital polarization $T_z$ (dashed lines) and exciton
    order parameter $P_1$ (solid line and symbols) as a function of the
    interaction strength $U$ at $J= 0.25 U$, left panel, and $J=0$. 
    For the $J=0$ case, we also show the order parameter $P_0$
    corresponding to the formation of an odd-parity spin-singlet
    excitonic state, which is always zero at $J=0.25 U$.}
  \label{fig1}
\end{figure}

To assess whether such excitonic phase indeed exists and does survive at intermediate coupling, we have calculated the dynamical 
susceptibility $\chi^{imp}_{11}(\omega) = \tfrac{1}{N}\int dt e^{i
  \omega t} \bra
T_t[\mathbb{\Gamma}_{11}(t)\mathbb{\Gamma}_{11}(0)]\ket$ forcing all
symmetries within the effective impurity problem of the
DMFT~\cite{GeorgesKotliar1996} and  
where $\mathbb{\Gamma}_{\alpha a} = \psi^\dag\Gamma_{\alpha
  a}\psi^\dagga$ \cite{Geffroy2019PRL,Blason2020PRB} are impurity
operators. 
Although
this quantity does not necessarily correspond to the local
susceptibility of the bulk model, nonetheless it provides suitable informations about its instabilities.   
In \figu{fig1}(a) we report the evolution of $\chi^{imp}_{11}(\omega)$, which is equivalent to $\chi^{imp}_{21}(\omega)$ by spin $U(1)$ symmetry, as a
function of energy $\omega$ and $U$ at $M=3.5$, thus along the path from 
the band to the Mott insulator. 
In the weakly interacting regime, this function displays several  
high energy peaks. Increasing $U$ leads to red shift of the
lowest energy peak until it softens before the
Mott transition sets in. The softening is just the signal of the excitonic instability.

However, the conclusive proof of excitonic transition can be obtained  
allowing for symmetry breaking, which we do though forcing, for
simplicity,  translational symmetry.
Our results are reported in \figu{fig1} for $J=0.25 U$, left panel, and $J=0$, right panel.

\subsection{The $J>0$ case}
For any $M>0$ we observe the formation of an EI with 
\bealn
P_1 = \langle\,\psi^\dagger_\ia\,
\Gamma_{11}\,\psi^\dagga_\ia\,\rangle\not= 0\,,
\eal
which is related to $\langle\,\psi^\dagger_\ia\,
\Gamma_{21}\,\psi^\dagga_\ia\,\rangle$ under spin $U(1)$, see Eq.~\eqn{order parameter}. 
The transition from the band or topological insulators to the excitonic one 
is of first order, while that from the EI to the high-spin Mott 
insulator (hs-MI) is of second order. We cannot exclude that also the latter transition may become first order allowing for translational symmetry breaking, and thus for an 
antiferromagnetic Mott insulator \cite{Amaricci2016PRB}.

The order parameter $P_1$ as a function of $U$ displays a
bell-like structure, which is centred at increasing values of $U$ as $M$ grows.
Interestingly, the peak value is attained at different positions
depending on the nature of the uncorrelated insulator. 
For $M<2$ (QSHI) the peak value is reached immediately after
the transition while for $M>2$ (BI) the peak is well inside
the excitonic region. We also observe that the orbital polarisation $T_z$ 
vanishes before the Mott transition, i.e., when $P_1$ is still finite.

\begin{figure}
  \includegraphics[width=0.5\textwidth]{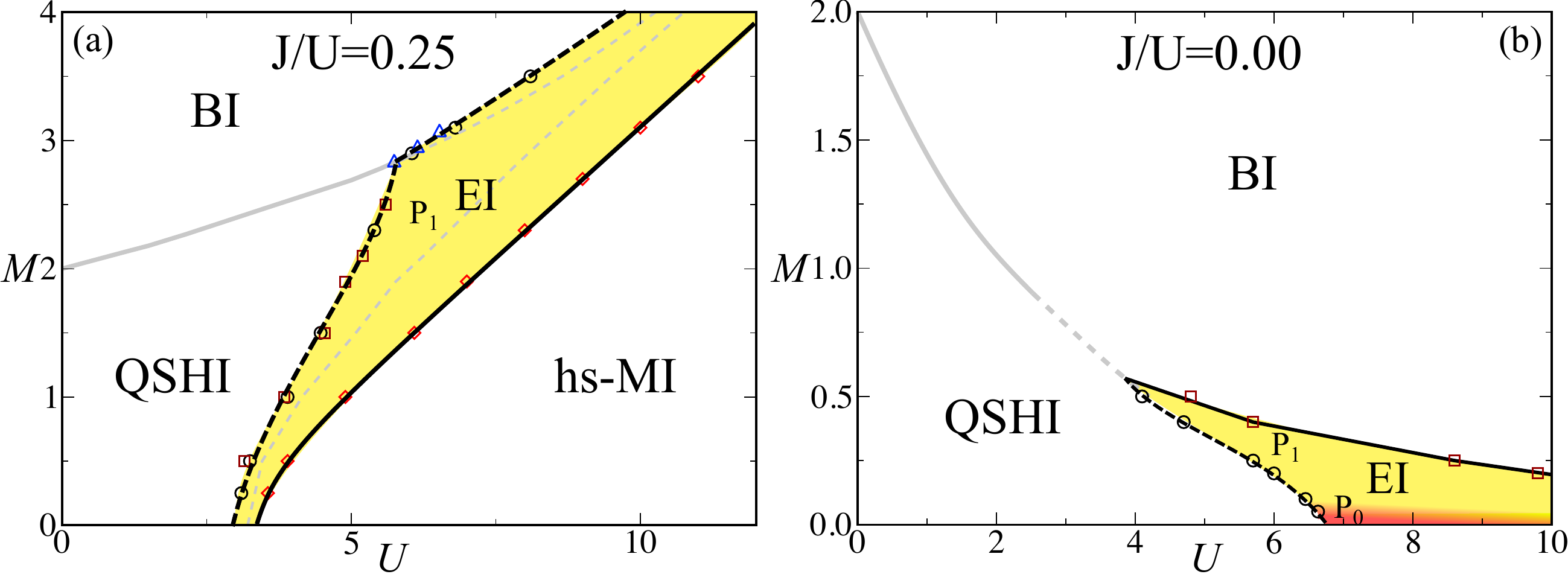}
  \caption{
    DMFT phase diagrams of the interacting model as a function of $U$ and
    $M$. Left panel (a) for $J/U=0.25$. Right panel (b) for
    $J=0$. The nature of the leading excitonic order parameter is
    indicated in the plot using text and color code. First order
    transition are indicated with dashed lines. Continuous
    transitions are indicated with solid lines. Transitions to/from
    EI are indicated in black. Gray lines in the background indicate
    the transitions occurring without allowing for exciton condensation. 
  }
  \label{fig3}
\end{figure}

\subsection{The $J=0$ case}
At $J=0$, the atomic levels \eqn{energies atomic limit} include the ground state $\mket{0,0,-1}$, followed at energy $2M$ above
by the fourfold multiplet $\mket{\pm 1,\pm 1,0}$ and $\mket{\pm 1,0,0}$, and, finally, by $\mket{0,0,+1}$ at energy $4M$ above the ground state. 
For large $U$, the hopping at second order in perturbation theory
generates superexchange processes of order $1/U$. 
Therefore, the model at $U\to \infty$ and finite $M$ describes just the band insulator with two electrons in orbital 2.  
However, the situation may change if $M$ scales as $1/U$. In that case, and if we 
discard the highest energy atomic level $\mket{0,0,+1}$, the superexchange processes mix the atomic ground state $\mket{0,0,-1}$ with the first excited 
multiplet on nearest neighbour sites. Similarly to the $J>0$ case, these processes  
may lead to finite expectation values of the local operators that have finite 
matrix elements between $\mket{0,0,-1}$ and the fourfold multiplet $\mket{\pm 1,\pm 1,0}\,\oplus\mket{\pm 1,0,0}$. We already showed that at $\lambda\not=0$ the 
mixing between $\mket{0,0,-1}$ and the doublet $\mket{+ 1,+ 1,0}\,\oplus\mket{- 1,- 1,0}$ stabilises the order parameters 
$P_1 = \langle\,\psi^\dagger_\ia\,\Gamma_{11}\,\psi^\dagga_\ia\,\rangle$ and its 
spin-$U(1)$ partner $\langle\,\psi^\dagger_\ia\,\Gamma_{21}\,\psi^\dagga_\ia\,\rangle$. \\
Similarly, the order parameters 
$P_0 = \langle\,\psi^\dagger_\ia\,\Gamma_{01}\,\psi^\dagga_\ia\,\rangle$ and its 
$C_4$ partner $\langle\,\psi^\dagger_\ia\,\Gamma_{32}\,\psi^\dagga_\ia\,\rangle$, 
which thus break $C_4$ and inversion symmetries, 
are favoured by the mixing between $\mket{0,0,-1}$ and the doublet 
$\mket{+ 1,0,0}\,\oplus\mket{- 1,0,0}$ at $\lambda\not=0$. \\
Our explicit DMFT calculations predict that, at $M\sim 1/U$, $P_1$ is always stabilised except at very small $M$, where the order parameter $P_0$ prevails, 
see right panel of \figu{fig1}. 
We further observe that at $J=0$ the transition from the QSHI to the EI is still 
first order, while that from the EI to the BI is continuous.

\begin{figure}
    \includegraphics[width=0.5\textwidth]{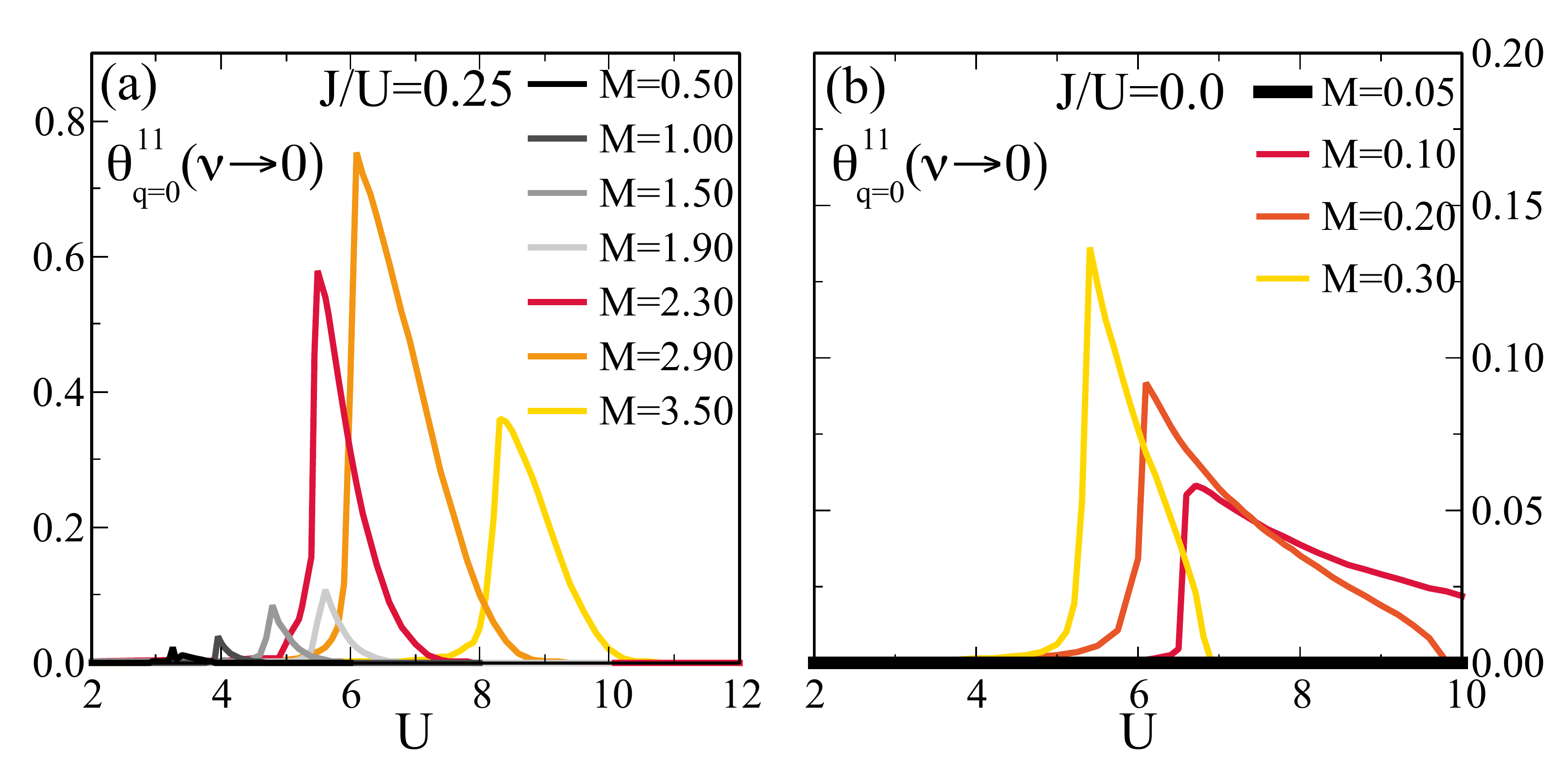}
    \caption{
      Static and uniform limit of the magnetic-electric susceptibility
      as a function of the interaction strength $U$ across the exciton phase
      transition. Data are for different values of $M$, as
      indicated in the panels, and for (a) $J/U=0.25$, (b) $J/U=0.00$. 
    }
  \label{fig4}
\end{figure}
\subsection{Phase diagrams}
We summarize our DMFT results in the two $U$ vs. $M$ phase diagrams 
at $J>0$ and $J=0$, respectively, left and right panels in \figu{fig3}. 
In both cases, the non-interacting QSHI-BI transition point at $M=2$ transforms at weak-coupling into a critical line determined by the
condition 
\bealn
M_\mathrm{eff}\equiv M+\tfrac{1}{4}\Tr{\big(\Gamma_{03}\Sigma(\omega=0)\big)}=2\,,
\eal
where $\Sigma(\omega)$ is the self-energy matrix.
The critical line corresponds to a second order phase transition up to a critical value of the interaction $U_c$. For $U>U_c$, the
transition turns first
order~\cite{Amaricci2015PRL,Mazza2016PRL,Roy2016},
thus without crossing a Dirac-like gapless point. 

For $J>0$ and large enough $U$, the ground state describes
a high-spin Mott insulator. An extended EI region with $P_1$ order parameter intrudes between the QSHI and the hs-MI, see \figu{fig3}(a).
Remarkably, the EI phase entirely covers the
discontinuous topological transition occurring between the BI and the
QSHI. The transitions from either the BI or the QSHI to the EI are of
first order, while
the transition from the EI to the hs-MI is
continuous. 

At $J=0$, we observe an EI region between the QSHI and the BI at small $M<0.5$. 
The QSHI-to-EI and EI-to-BI transitions are, respectively, of first and second orders. For very large $U$, the EI phase appears at $M$ scaling as $1/U$. 
As we mentioned, the exciton condensate with order parameter $P_0$, breaking 
$C_4$ and inversion symmetries, for very small $M$, while
for larger values, the order parameter $P_1$ prevails, 
breaking inversion, time-reversal and spin $U(1)$. The transition between 
$P_1$ and $P_0$ is expected to be first order.

\begin{figure}
  \includegraphics[width=0.5\textwidth]{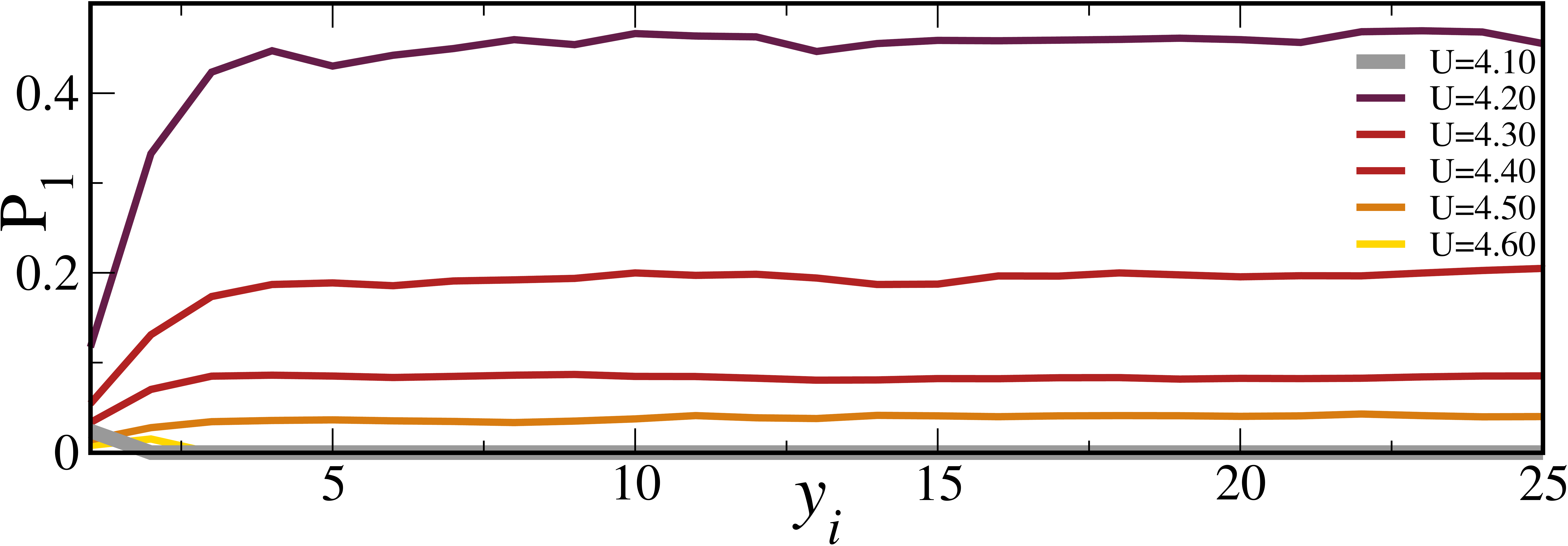}
  \includegraphics[width=0.5\textwidth]{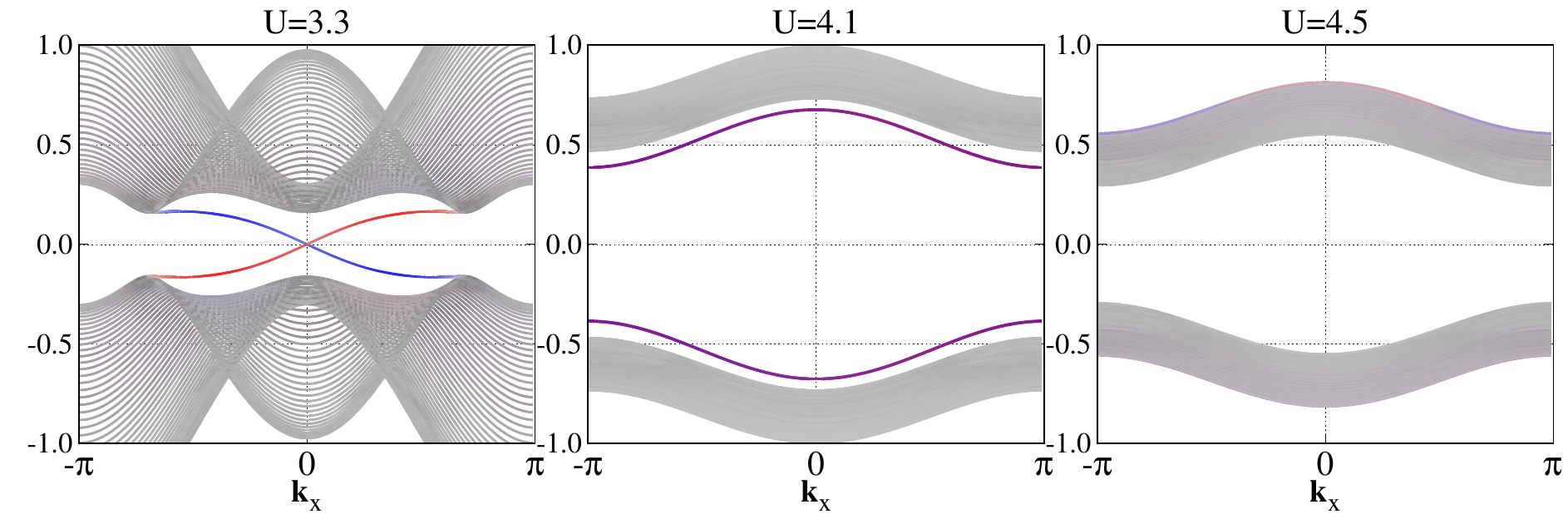}
  \caption{
    Top panel:
    Bottom panels:
    Evolution of the low energy band structure of the
    interacting model at $M=1$ and $J=0.25 U$ on a slab geometry 
    across the QSHI to EI transition. 
  }
  \label{fig5}
\end{figure}
\section{Magneto-electric nature of the excitonic insulator}\label{Sec3}
In the EI phase with $P_1$ order parameter, the breakdown
of the symmetries protecting the non-trivial topology of the QSHI,
i.e., time-reversal $\TT$, inversion $\PP$
and spin $U(1)$ (yet not the product $\PP\TT$), dramatically changes
the response to an electromagnetic field.
Specifically, the triplet in-plane spin polarization nature of the
excitonic order parameter forbids a direct coupling to 
the electric field and, independently, to the magnetic field. 
However, the lack of both $\TT$ and $\PP$ symmetries allows the system to
couple to the product of magnetic and electric field, i.e. a linear
magneto-electric (ME) response~\cite{Eerenstein2006N}.   
Here, we show that the EI admits a finite
ME susceptibility and thus corresponds to a ME insulator.
Notably, this state should not be expected to be multi-ferroic because
of the absence of magnetic and electric order~\cite{Eerenstein2006N}.

In order to study the ME properties, we evaluate the
electric dipole response to a magnetic perturbation. Using Green-Kubo
formalism and neglecting vertex corrections, we obtain the following expression:  
\begin{equation}
  \begin{split}
    &\Theta^{ab}_\qa(\nu_m) = \\
    &\sum_{\ka,n} \frac{\Tr\left[G(\ka,\iome) p^a(\ka) G(\ka\!+\!\qa,\iome\!+\!i\nu_m)M^b \right]}{\beta\nu_m}
  \end{split}
\end{equation}
where $a,b=1,2\==x,y$ are the in-plane directions, $M^a=\tfrac{1}{2}\Gamma_{a0}$
is the $a$ component of the spin operator, $p^a(\ka)$ is the
momentum operator along $a$, $\iome$ and $i\nu_m$ are,
respectively, fermionic and bosonic Matsubara frequencies and 
\bealn
G(\ka,\iome) = \Big(\iome +\mu - H(\ka) - \Sigma(\iome)\Big)^{-1}\,,
\eal
is the interacting Green's function matrix. 
Given the multi-orbital nature of the Hamiltonian (\ref{Hbhz}) the momentum operator
should be evaluated using a generalized Peierls
approximation
\cite{Cruz1999PRB,Pedersen2001PRB,Tomczak2009PRB,Wissgott2012PRB,Mazza2019PRL,Li2020PRB}.
The latter includes additional contributions, stemming from 
on-site inter-orbital
processes that are dipole allowed. Specifically,    
$$
p^a_{\a\b}(\ka) = \partial^a H_{\a\b}(\ka) + i\Omega_{\a\b}(\ka)d^a
$$ 
where $\Omega^{\a\b}(\ka)=[E_\a(\ka)-E_\b(\ka)]$, $E_\a(\ka)$ are the
eigenvalues of the non-interacting Hamiltonian (\ref{Hbhz}), and
$\vec{d}=\left(\Gamma_{02},\Gamma_{32}\right)$ is the dipole operator.

In the following we consider the static, $\nu\rightarrow 0$, and uniform,
$\qa=0$, limit of the ME susceptibility.  
Our results are presented in \figu{fig4}, where we shows the evolution of $
\Theta^{11}_{\qa=0}(\nu=0)$ as a function of $U$ for finite and zero
values of $J$. Since the contribution of the group velocity
$\partial_{k_a}H_{\a\b}(\ka)$ vanishes by symmetry, the ME response is
entirely determined by the intra-atomic dipole transitions, which have finite 
expectation values in the EI phase.   
Indeed, $\Theta^{11}_{\qa=0}$ is finite only within the EI phase
and vanishes otherwise.
The magneto-electric susceptibility shows the same dome structure 
of the order parameter $P_1$ as a function of $U$. 
The results at $J=0$ reported in \figu{fig4}(b) point out that the ME response 
vanishes when $P_0\not =0$, as expected by symmetry. 
In the EI phase with $P_1\not=0$, the ME
susceptibility is finite and its peak value shifts to lower $U$ with increasing $M$.\\  
At $J>0$, see \figu{fig4}(a), 
we observe a substantial change in the magnitude of the ME
response. For $M<2$, thus starting from the QSHI, the
susceptibility is globally small, while for larger $M$ the
weight of $\Theta^{11}_{\qa=0}$ increases, with 
seven times larger peak values.    
Remarkably, for any given $J$ the largest ME response is reached in
proximity of the quantum critical point which, without allowing for $P_1\not=0$, separates the continuous from the first order topological
quantum phase transition \cite{Amaricci2015PRL,Amaricci2016PRB}.

\section{Slab geometry and edge states}\label{Sec4}
Finally, we explore the evolution across the QSHI-to-EI 
phase transition at $J>0$ in a slab geometry,
i.e. with open boundary conditions along, say, the $y$ axis, and periodic 
in the perpendicular direction. 
In this geometry, the electrons at the boundary
experience an effectively larger interaction strength because of
the reduced coordination. This effect becomes detectable
near the phase transition.
In the top panel of \figu{fig5}, we show the evolution of the
$P_1$ order parameter across the QSHI-EI first order transition with $M=1$.
Before the transition, a finite value of the order parameter appears at the
boundary, and fast decays in the bulk interior. This is akin a wetting phenomenon that arise since the more correlated surface favours the nucleation of the 
EI phase near the first order QSHI-EI transition. 
This result is remarkable, since it predicts that a QSHI might display 
a surface layer of excitonic insulator without topological edge states
but with finite magnetoelectric response.  
Increasing $U$ above the value of the bulk transition, drives the sudden formation of a finite order parameter throughout the sample, as expected for a first order 
transition. In this case, the order parameter near the boundary is instead 
reduced with respect to the bulk, consistently with the behaviour of $P_1$ versus 
$U$ for $M<2$, see left panel in \figu{fig1}.

Further insights can be gained investigating the fate of the
helical edge states, see bottom panels of
\figu{fig5}.
The plots show the low-energy electronic band structure of the
interacting system across the QSHI to EI transition.
At small coupling $U=3.3$, well inside the QSHI (left panel), gapless helical edge states well separated from the bulk spectrum are visible. However, at $U=4.1$ (middle panel), 
where the bulk is still a QSHI but an EI wetting layer has formed, see top panel 
of \figu{fig5}, edge states still exist but are gapped. 
On the contrary, for $U=4.5$ (right panel), where also the bulk is an EI, the edge states have disappeared inside the bulk continuum.    

\section{Conclusions}\label{Sec5}
In conclusions, we have investigated a canonical model of 
interacting quantum spin Hall insulators and showed that for a strong
enough electronic correlation the system gets generally unstable towards an
excitonic insulator that breaks time-reversal and inversion
symmetries, as well as the residual spin $U(1)$ rotations. This state
further evolves into a magnetic Mott insulator upon increasing the
interaction strength, where inversion
and spin-$U(1)$ symmetry are recovered. We explicitly show that the 
excitonic insulator has non-zero magneto-electric
susceptibility, and thus is a good candidate platform for the
realization of correlated multi-ferroic materials.
Another remarkable phenomenon that we uncovered is the possible
existence of an excitonic insulator wetting layer in a quantum spin Hall insulator.

\section{Acknowledgements}
A.A., M.C. and M.F. acknowledge support from H2020 Framework Programme, 
under ERC Advanced Grant No. 692670 FIRSTORM. A.A. and M.C. also
acknowledge support from Italian MIUR through PRIN2017 CEnTral
(Protocol Number 20172H2SC4).
G.M. was supported by the Swiss FNS/SNF through an Ambizione grant.

\bibliography{references}

\end{document}